\documentclass[prb,
twocolumn,
superscriptaddress,showpacs,amsmath,amssymb]{revtex4}
\usepackage{amsfonts}
\usepackage{bm}
\usepackage{verbatim}

\usepackage{graphicx}

\begin{document}

\title{Comment to the CPT-symmetric Universe: Two possible extensions.}

\author{G.E.~Volovik}
\affiliation{Low Temperature Laboratory, Aalto University,  P.O. Box 15100, FI-00076 Aalto, Finland}
\affiliation{Landau Institute for Theoretical Physics, acad. Semyonov av., 1a, 142432,
Chernogolovka, Russia}

\date{\today}

\begin{abstract}
In Ref. \cite{Turok2018} the antispacetime Universe was suggested as the analytic continuation of our Universe across the Big Bang singularity in conformal time. We consider two different scenarios of analytic continuation. In one of them the analytic continuation is extended to the temperature of the system. This extension suggests that if such analytic continuation is valid, then it is possible that
the initial stage of the evolution of the Universe on our side of the Big Bang was characterized by the negative temperature. In the second scenario, the analytic continuation is considered in the proper time. In this scenario the Big Bang represents the bifurcation point at which the $Z_2$ symmetry between the spacetime and antispacetime is spontaneously broken.
 \end{abstract}
\pacs{
}

\maketitle

\section{Introduction}

Recently there was suggestion to extend the Universe beyond the Big Bang using the analytic continuation of the radiation-dominated epoch across the singularity\cite{Turok2018,Turok2018b}. In this analytic continuation, at which the scale factor $a(\tau)$ changes sign at $\tau=0$, the gravitational tetrads also change sign giving rise to what is called the antispacetime. This means that the Universe on the other side of the Big Bang is the mirror image of the Universe on our side of the Big Bang.

Different types of the antispacetime obtained by the space reversal $P$ and time reversal $T$ operations  were considered ealier, including those where the determinant of the tetrads $e$ changes sign.\cite{Diakonov2011,Diakonov2012,Rovelli2012b,Rovelli2012a,NissinenVolovik2018,Volovik2019b,Vergeles2019}
Later the consideration has been extended to the thermal states, where the possible analytic continuation of the temperature across the transition from spacetime to antispacetime has been considered \cite{Volovik2019}. Here we discuss two different scenarios of analytic contunuation across the Big Bang.

\section{CPT-symmetric Universe and negative temperature}

The CPT-symmetric Universe has been obtained in the conformal time frame\cite{Turok2018}, where the metric in the spatially flat radiation-dominated era is:
\begin{equation}
g_{\mu\nu}=a^2(\tau)\eta_{\mu\nu}\,,
\label{metric}
\end{equation}
where $\eta_{\mu\nu}$ is the flat Minkowski metric; $a(\tau)$ is the scale factor and $\tau$ is the conformal time. 
Since the scale factor $a(\tau)\propto\tau$, it was suggested that it can be analytically transformed to the region $\tau<0$ before the Big Bang. Then the terad fields, which are proporional to  $a(\tau)$, also change sign under this analytic continuation,
$e^a_\mu(-\tau)=-e^a_\mu(\tau)$. 

Now let us go further and extend the analytical continuation to the temperature of the system.
In the spatially flat radiation-dominated era one has
\begin{equation}
T(\tau)\propto \frac{1}{N(\tau)}= \frac{1}{e_{00}(\tau)}  \,\,, {\rm or} \,\, \beta(\tau)\propto\tau\,.
\label{TinCPT}
\end{equation}
The analytic continuation of $\beta(\tau)$ suggests that in the Universe at $\tau<0$ the temperature is negative.
The transition between the states with positive and negative temperatures occurs via the infinite $T$ at $\tau=0$ ($\beta(0)=0$).
The states with negative temperature are typically unstable thermodynamically. This means that if the analytic continuation is really valid, then one of the two states (before or after Big Bang) is thermodynamically unstable.
It is more natural to assume, that the evolution of the system from $\tau = -\infty$ to $\tau=0$ was equilibrium and corresponded to the positive temperature, $T(\tau<0)>0$. This is because the system had enough time to equilibrate before the collapse. So we should have the following analytic time dependence of temperature:
\begin{equation}
T(\tau)\propto -\frac{1}{e_{00}(\tau)} \,\,, {\rm or} \,\, \beta(\tau)\propto -\tau\,,
\label{TinCPTopposite}
\end{equation}
and thus on our side of the Big Bang the temperature is negative, $T(\tau>0)<0$. Of course, this may happen only at the first stage of the evolution of our Universe, i.e. immediately after the Big Bang. After some time the equilibration occurs, and the system returns again to the evolution with the equilibrium positive temperature,  $T(\tau>0)>0$.
The proper justification of this suggestion is beyond this comment.

\section{Euclidean signature Universe and spontaneously broken CPT symmetry}

%%%%%%%%%%%%%%%%%%%%%%%%%%%%%%%%%%%%%%%%%%%%%%%%%%%%%%%%%
%%%%%%%%%%%%%%%%%%%%%%%%%%%%%%%%%%%%%%%%%%%%%%%%%%%%%%%%%
\begin{figure}[t]
\includegraphics[width=\linewidth]{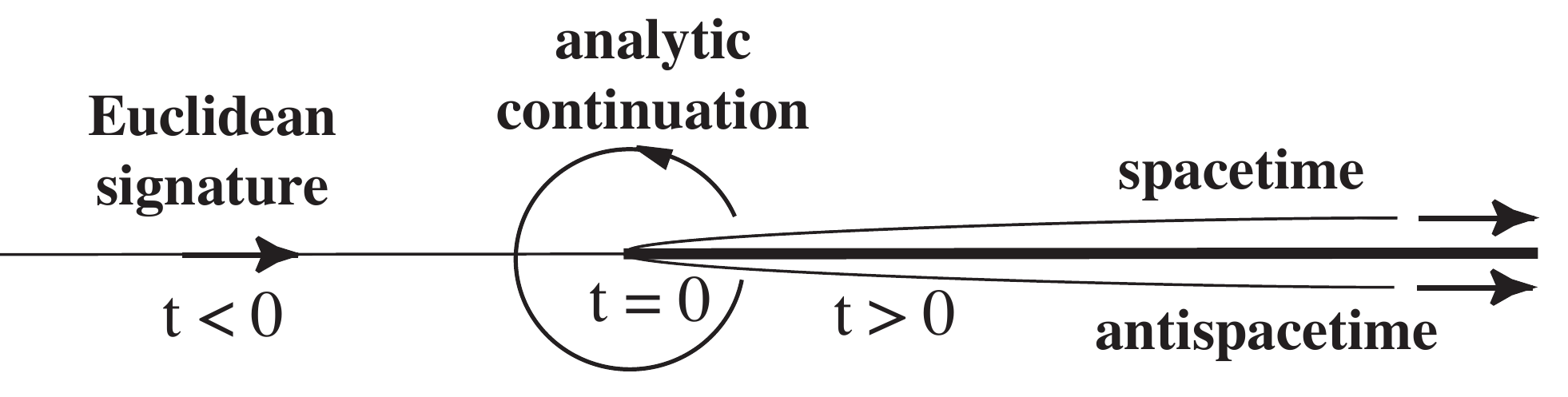}
\caption{ Bifurcation at Big Bang. Analytic continuation in physical proper time $t$: the scale factor
$a(t)\propto \sqrt{t}$ changes sign around the Big Bang point, transforming spacetime into antispacetime.
At $t<0$ the scale factor is imaginary, which corresponds to the metric with Euclidean  signature.
Crossing the Big Bang from the $t<0$ semiaxis, the Universe approaches either spacetime or antispacetime. In this scenario, the Big Bang represents the bifurcation point at which the $Z_2$ symmetry 
between the spacetime and antispacetime is spontaneously broken. The quantum superposition of the two states is not allowed  in the macroscopic system.
}
\label{TwoRoads_Fig}
\end{figure}
%%%%%%%%%%%%%%%%%%%%%%%%%%%%%%%%%%%%%%%%%%%%%%%%%%%%%%%%%
%%%%%%%%%%%%%%%%%%%%%%%%%%%%%%%%%%%%%%%%%%%%%%%%%%%%%%%%%

Let us consider the analytic continuation in terms of the physical proper time $t$. The metric for the radiation-dominated Universe is:
\begin{equation}
g_{\mu\nu}=dt^2 -a^2(t)d{\bf r}^2\,\,,\, a^2(t)  \propto t \,.
\label{metric2}
\end{equation}
let us now assume that the scale factor $a(t) \sim \sqrt{t}$ can be analytically extended around the singularity at $t=0$. Then the spacetime (with positive spatial components of tetrads) and antispacetime
 (with negative spatial components of tetrads) can be connected by analytic continuation: by $2\pi$ rotation about $t=0$. As distinct from scenario in \cite{Turok2018}, where the two Univereses live together,
in the proper time scenario the spacetime and antispacetime represent two different realizations of the Universe at $t>0$, which exclude each other. 

The Universe at $t>0$ can be obtained by analytic continuation from the negative $t$ region, where the metric has  Euclidean signature. In this scenario, the Big Bang (at $t=0$) represents the point of 
the phase transition from the Euclidean to Minkowski signature, which is similar to that, say,  in \cite{Steinacker2018,Stern2018,Sakharov1991,GibbonsHartle1990} (an example of such transition in condensed matter can be found e.g. in Ref. \cite{NissinenVolovik2017b}). At the Big Bang, the state with Euclidean signature transforms either to the spacetime or to antispacetime, but not to the quantum superposition of the two Universes. The latter is not allowed for the macroscopic systems, which experience the spontaneous symmetry breaking.\cite{Grady1994}
In principle, immediately after the Big Bang the Universe may evolve as the quantum superposition of these two states, but due to rapid decoherence only one of the two states survives. Thus in this analytic continuation the $Z_2$ ($CPT$) symmetry is spontaneously broken, as distinct from the scenario in \cite{Turok2018}, which
describes the creation of two Universes both evolving to the future with the conservation of the  $CPT$ symmetry.

\section{Discussion}

In conclusion, we extended the analytic continuation across the Big Bang proposed in Ref.\cite{Turok2018}in two different ways. The analytic continuation in the conformal time frame $\tau$ was extended to the temperature of the system. This extension suggests that if the analytic continuation of the terad is valid, then it is possible that
the initial stage of the evolution of our Universe after the Big Bang was characterized by the negative temperature. The analytic continuation in the proper time $t$  suggests that the  Big Bang is the bifurcation point of the second order quantum transition from the Euclidean to Minkowski signature, at which the symmetry between the spacetime and antispacetime is spontaneously broken.

{\bf Acknowledgements}. 
 I thank A. Starobinsky for criticism. This work has been supported by the European Research Council (ERC) under the European Union's Horizon 2020 research and innovation programme (Grant Agreement No. 694248).


\begin{thebibliography}{99}

\bibitem{Turok2018}
L. Boyle, K. Finn and N. Turok,
CPT-Symmetric Universe, 
Phys. Rev. Lett. {\bf 121}, 251301 (2018).

\bibitem{Turok2018b}
L. Boyle, K. Finn and N. Turok,
The Big Bang, CPT, and neutrino dark matter,
arXiv:1803.08930.

\bibitem{Diakonov2011}
D. Diakonov,
Towards lattice-regularized Quantum Gravity,
  arXiv:1109.0091.

\bibitem{Diakonov2012}
A.A. Vladimirov, D. Diakonov,
Phase transitions in spinor quantum gravity on a lattice,
Phys. Rev. D {\bf 86}, 104019 (2012). 

\bibitem{Rovelli2012b}
M. Christodoulou, A. Riello, C. Rovelli,
How to detect an anti-spacetime,
Int. J. Mod. Phys. D {\bf 21}, 1242014 (2012),
arXiv:1206.3903.

\bibitem{Rovelli2012a}
C. Rovelli, E. Wilson-Ewing,
 Discrete symmetries in covariant LQG,
Phys. Rev. D {\bf 86}, 064002 (2012),
 arXiv:1205.0733.

\bibitem{NissinenVolovik2018}
J. Nissinen and G.E. Volovik,
Dimensional crossover of effective orbital dynamics in polar distorted  $^3$He-A: Transitions to anti-spacetime,
Phys. Rev. D {\bf 97}, 025018  (2018),

\bibitem{Volovik2019b}
G.E. Volovik,
Two roads to antispacetime in polar distorted B phase: Kibble wall and half-quantum
vortex,
arXiv:1903.02418.

\bibitem{Vergeles2019}
S.N. Vergeles
A note on the vacuum structure to lattice Euclidean quantum gravity,
arXiv:1903.09957.

\bibitem{Volovik2019}
G.E. Volovik,
Negative temperature for negative lapse function,
Pis'ma ZhETF  {\bf 109}, 10-11 (2019),
arXiv:1806.06554.


\bibitem{Steinacker2018}
H.C. Steinacker,
Cosmological space-times with resolved Big Bang in Yang-Mills matrix models,
High Energ. Phys. 02  (2018)  033,
arXiv:1709.10480.


\bibitem{Stern2018}
A. Stern and Chuang Xu,
Signature change in matrix model solutions,
Phys. Rev. D {\bf 98}, 086015 (2018).

\bibitem{Sakharov1991}
A.D. Sakharov,
Cosmological transitions with changes in the signature of the metric,
Sov. Phys. Usp. {\bf 34}, 409--413 (1991).

\bibitem{GibbonsHartle1990}
G. W. Gibbons and J. B. Hartle,
Real tunneling geometries and the large-scale topology of the universe,
Phys. Rev. D {\bf 42}, 2458 (1990) 

\bibitem{NissinenVolovik2017b}
J. Nissinen and G.E. Volovik,
Effective Minkowski-to-Euclidean signature change of the magnon BEC pseudo-Goldstone mode in polar $^3$He,
Pis'ma ZhETF {\bf 106}, 220--221 (2017), 
JETP Lett. {\bf 106},  234--241 (2017),
 arXiv:1707.00905.

\bibitem{Grady1994}
M. Grady,
Spontaneous symmetry breaking as the mechanism of quantum measurement,
hep-th/9409049.

\end{thebibliography}
\end{document}